\documentclass{appolb}
\usepackage[latin1]{inputenc}
\usepackage{graphicx}
\usepackage{amsmath}
\newcommand{\De}[3][]{\frac{\partial^{#1} #2}{\partial #3^{#1}}}
\newcommand{\eps}{\varepsilon}
\newcommand{\s}{\boldsymbol{s}}
\DeclareGraphicsExtensions{.pdf}
\begin{document}
\pagestyle{plain}
\newcount\eLiNe\eLiNe=\inputlineno\advance\eLiNe by -1
\title{Smallworld bifurcations in an opinion model
\thanks{Presented at the Summer Solstice 2009 International Conference
on Discrete Models of Complex Systems
Gdansk, Poland, 22-24 June 2009.}}%

\author{  Franco Bagnoli
  \address{Dept. Energy and CSDC University of Florence, via S. Marta, 3 I-50139 Firenze, Italy. Also  INFN, sez. Firenze. email:franco.bagnoli@unifi.it}
  \and Graziano Barnabei
  \address{Dept. Systems and Computer Science and CSDC, University of Florence.}
  \and Ra\'ul Rechtman
  \address{Centro de Investigaci\'on en Energ\'i{}a, Universidad Nacional Aut\'onoma de M\'exico, Temixco Mor., M\'exico}
}
\maketitle

\begin{abstract}
We study a cellular automaton opinion formation model of Ising type, with antiferromagnetic pair interactions modeling anticonformism, and ferromagnetic plaquette terms modeling the social norm constraints.
For a sufficiently large connectivity, the mean-field equation for the average magnetization (opinion density) is chaotic. This ``chaoticity'' would imply irregular coherent oscillations of the whole society, that may eventually lead to a sudden jump into an absorbing state, if present.

However, simulations on regular one-dimensional lattices show a different scenario: local patches may oscillate following the mean-field description, but these oscillations are not correlated spatially, so the average magnetization fluctuates around zero (average opinion near one half). The system is chaotic, but in a microscopic sense where local fluctuations tend to compensate each other.

By varying the long-range rewiring of links, we trigger a small-world effect. We observe a bifurcation diagram for the magnetization, with period doubling cascades ending in a chaotic phase. As far as we know, this is the first observation of a small-world induced bifurcation diagram.

The social implications of this transition are also interesting. In the presence of strong ``anticonformistic'' (or ``antinorm'') behavior, efforts for promoting social homogenization may trigger violent oscillations.
\end{abstract}

\section{Introduction}

Social norms are the basis of a community. Social norms are often adopted and respected even if in contrast with an individual's immediate advantage, or, alternatively, even if they are costly with respect to a ``naive'' behavior. Indeed, the social pressure towards a widespread social norm is sometimes  more powerful than a norm imposed by punishments.

On the other hand, it is well known that the establishment of social norms is difficult to plan, and their imposition is hard to be fulfilled. This problem has been affronted by Axelrod in a game-theoretic formulation~\cite{Axelrod:EvolutionOfCooperation}, as the foundation of the cooperation and of the society itself. Axelrod's idea is that of a repeated game. Although in an one-shot game it is always profitable to win not following any norm, in a repeated game there might be several reasons for cooperation~\cite{Nowak:FiveRules}, the most common ones are direct reciprocity and reputation. In all these games, the crucial parameters are the cost for cooperation with respect to defeat, and the expected number of re-encounters with one's opponent or the probability that one's behavior will become public. One can assume that these aspects are related to the size of the local community with which one interacts, and on the fraction of people in this community that share the acceptance of the social norm. Indeed, the behavior of a spatial social game is strongly influenced by the network structure~\cite{Klemm}.

The previous approach assumes perfect rationality of agents, and does not take into consideration ``irrational'' tendencies like for instance education. It is  well known that a given predisposition towards conformism or anticonformism (\emph{i.e.}, the education by parents, school and the social community) may influence the acceptance of a given social norm.

In this paper we model the dynamics of the acceptance of a social norm in a community with different degrees of conformism or anticonformism. We shall consider a simplified cellular automata model, already introduced in Ref.~\cite{Bagnoli:ChaosUniformSociety}. See Ref.~\cite{review} for a detailed review of models of social dynamics.

\section{Cellular automata model}

Let us denote by $s_i= s_i(t)$ the opinion of individual $i$ at time $t$. We consider the case of two opinions in competition $s_i\in \{0,1\}$. We can switch to Ising-like variables (spin) $\sigma_i\in\{-1,1\}$ by the transformation
\[
 \sigma_i = 2s_i-1.
\]

The individual opinion evolves in time according to the opinions of  neighbors, identified by an adjacency matrix $a_{i,j}\in\{0,1\}$. This matrix defines the network of interactions and is considered fixed in time. An individual may be part of his neighborhood.

The neighborhood of an individual $i$ is the set of individuals $j$ such that $a_{i,j}=1$.
The connectivity $k_i$ of individual $i$ is the number of nonzero entries in $a_{i,j}$, \emph{i.e.},
\[
 k_i = \sum_j a_{i,j}.
\]

In the following, we shall consider uniform neighborhoods with connectivity $k$, where the actual neighborhood can be either regular 
\[
 a_{i,j} = \begin{cases}
           1 &\text{if $j-i < k$,} \\
           0 & \text{otherwise.}
          \end{cases}
\]
or partially rewired, where a fraction $p$ of the nonzero entries of each row of $a_{i,j}$ is set to zero, and replaced with a randomly chosen $j'$ (with $a_{i,j'}=0$), setting $a_{i,j'}=1$.

We choose to assign equal weight to all neighbors, so we define the local field (social pressure) $h_i$ as
\begin{equation}\label{h}
 h_i =\dfrac{\sum_{j} a_{i,j}s_j}{k_i}.
\end{equation}
The local field takes values between 0 and 1. We might also add an external field $H$, modeling written or broadcasting media, but in this study we always keep $H=0$.

We are modelling here a completely uniform society, \emph{i.e.}, we assume that the individual variations in the response to stumuli are quite small. Moreover, we do not include any memory effect, so that the dynamics is completely Markovian.

The effects of the social pressure is assumed to be proportional to the social field.   The evolution is defined by the transition probabilities
\[
  \tau(1|h_i),
\]
denoting the probability of observing a spin $s_i(t+1)=1$ given a local field $h_i$ at time $t$. Clearly, $ \tau(0|h_i) = 1-  \tau(1|h_i)$.

By denoting by $\s$ a spin configuration $(s_1, s_2, \dots)$, and by $P(\s,t)$ the probability of observing it at time $t$, we have
\begin{equation}\label{Markov}
 P(\s',t+1) = \sum_{\s} W(\s'|\s) P(\s,t)
\end{equation}
with
\[
 W(\s'|\s) = \prod_i \tau(s'_i|h_i),
\]
and $h_i$ given by Eq.~\eqref{h}.

If no transition probability is zero or one, we can map the Markov matrix $W$ onto a dynamic equilibrium model, of Ising type~\cite{DomanyKinzelPRL}.

Eq.~\eqref{Markov} may be equivalently written using restricted distributions.
Let us call $p_n(s_i,\dots,s_{i+n}; t)$ the (restricted) probability of observing the sequence $s_{i}, \dots,s_{i+n}$ at time $t$, defined as
\[
	p_n(s_i,\dots,s_{i+n}; t) = \sum_{s_j : j \notin \{i,\dots,i+n\}} P(s_1,\dots,s_i,\dots,s_{i+n}, \dots;t).
\]
By assuming spatial uniformity, the $p_n$ do not depend on the index $i$. The quantity  $p_1(1,t)$  is the usual density, that will be denoted by $c$.

Let us consider as an example the one-dimensional regular lattice case with uniform connectivity $k=2$, $a_{i,i}=a_{i,i+1} =1$. Using the restricted distributions, the evolution equation of the system is given by the infinite hierarchy
\begin{equation}\label{hier}
\begin{split}
  p_1(s'_1;t+1) &= \sum_{s_1, s_2} p_2(s_1, s_2; t) \tau(s'_1|s_1,s_2)\\
  p_2(s'_1,s'_2;t+1) &= \sum_{s_1, s_2,s_3} p_3(s_1, s_2, s_3; t) \tau(s'_1|s_1, s_2)\tau(s'2|s_2, s_3)\\
  p_3(s'_1,s'_2, s'_3;t+1) &=\dots
\end{split}
\end{equation}
where for readability (and generality)  we have written $\tau(s'_1|s_1,s_2)$ instead of $\tau(s'_1|(s_1+s_2)/2)$.

\section{Phase transitions in uniform societies}

\begin{figure}
 \begin{center}
  \includegraphics[width=0.7\textwidth]{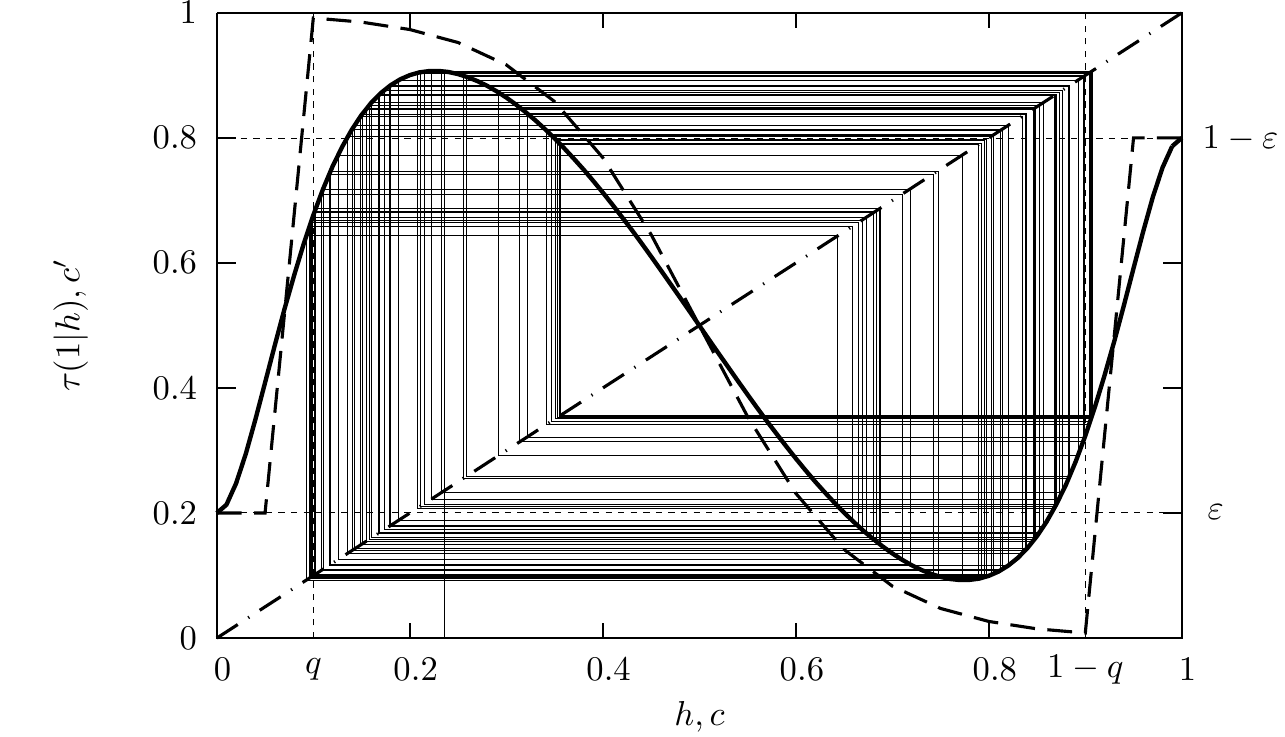}
  \caption{\label{fig:mf} Transition probabilities $\tau(1|h)$ (Eq.~\eqref{tau}, thick dashed  line) and return map $f(c)$ (Eq.~\eqref{mf}, thick continuous line) as a function of $c$. Some iterations of the map are also shown (thin continuous line) and the bisectrix (dash-dotted line). The threshold values $q$ and $1-q$ are marked by thin dashed vertical lines, and the values $\eps$ and $1-\eps$ by thin dashed horizontal lines. Here $k=20$, $q=0.1$, $J=-3$, $\eps = 0.2$. }
 \end{center}
\end{figure}

Let us express the transition probabilities as
\begin{equation}\label{tau}
 \tau(1|h) =
  \begin{cases}
    \eps & \text{if $h<q$,}\\
    \dfrac{1}{1+\exp(-2J(2h-1))} & \text{if $q\le h\le 1-q$,}\\
    1-\eps & \text{if $h>1-q$,}
  \end{cases}
\end{equation}
shown in Fig.~\ref{fig:mf}.
In this way we model a standard dynamic Ising model with ferromagnetic (for $J>0$) or antiferromagnetic ($J<0$) interactions, with plaquette terms given by $\eps$\footnote{A similar model can be defined using a standard Hamiltonian formalism by including ferromagnetic plaquette terms proportional to $(2h-1)^3$ or higher odd powers.}. For $\eps=0$ (infinite plaquette terms) we have the absorbing states $\s=0$ and $\s=1$ if the social pressure is above (below) the threshold  $1-q$ ($q$), respectively.

In one dimension, with $k=3$, $1/3<q\le 1/2$ and $\eps=0$ this model exhibits a nontrivial phase diagram~\cite{Bagnoli:3inputs}, with two directed-percolation transition lines that meet a first-order transition line in a tricritical point, belonging to the parity conservation universality class. Essentially, we have the stability of the two absorbing states for $J>0$ (ferromagnetic interactions, conformistic society, ordered phase), while for $J<0$ (anti-ferro and anti-conformistic) the absorbing states are unstable and a new, disordered active phase is observed. This scenario corresponds essentially to the simplest mean-field picture. The interpretation of mean-field predictions is that $0<c<1$ corresponds to the active phase, which is microscopically \emph{chaotic}, with the appearance of transient correlated patches (``triangles''). This is due to the presence of the unstable absorbing states: occasionally a patch of sites ``fall'' into one of these states. Since the absorbing state is absorbing, it can be abandoned only by ``erosion'' at boundaries, and this originates the  ``triangular'' pattern.

The model has been studied also in the one-dimensional case with larger neighborhood~\cite{Bagnoli:longrange}. In this case we observe again the transition from an ordered to an active, microscopically chaotic phase, but this transition occurs through a disordered phase, with no apparent structure in the time-space pattern, which is moreover insensitive to variations of $J$. Indeed, if the system ``falls'' into a truly disordered configuration, then $h$ is everywhere equal to $0.5$ and  the transition probabilities $\tau$ become insensitive to $J$ and equal to $0.5$, Eq.~\eqref{tau}. This \emph{disordered} regime is therefore different from the ``microscopically chaotic'' one.

\begin{figure}
 \begin{center}
  \includegraphics[width=0.7\textwidth]{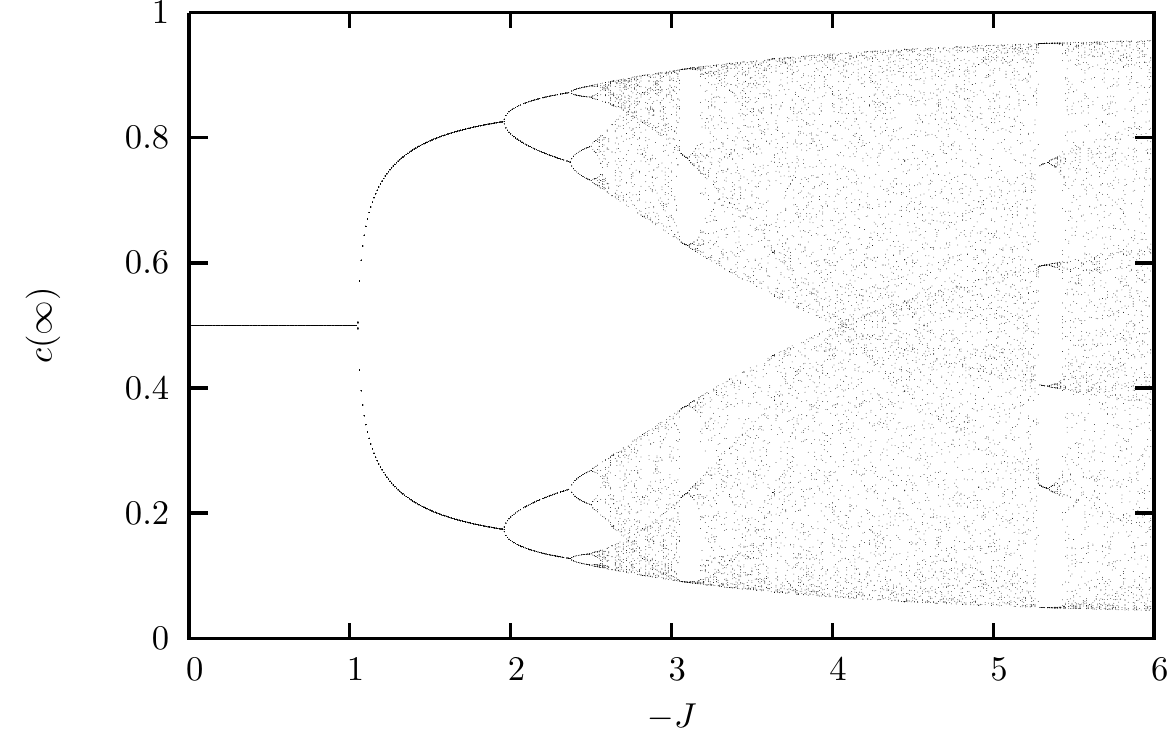}
  \caption{\label{fig:mfb} Mean field bifurcation diagram of Eq.~\eqref{mf} for $0\le -J \le 6$ and  $k=20$, $q=0.1$, $\eps = 0.2$. Transient of 100 time steps, and plot of 10 iterations for 4 random initial conditions.}
 \end{center}
\end{figure}

\section{Mean-field chaotic behavior}

The mean field approximation consists in truncating the previous hierarchy, Eq.~\eqref{hier}, at a given point by factorizing the distribution probabilities. At the lowest level, for a uniform connectivity $k$, one gets
\begin{equation}\label{mf}
  c' = f(c)=\sum_{v=0}^k \binom{k}{v} c^v (1-c)^{k-v} \tau(1| v/k),
\end{equation}
which is also shown in Fig.~\ref{fig:mf}.
In this approximation, the return map Eq.~\eqref{mf} may become chaotic, for antiferromagnetic ($J<0$) couplings and sufficiently large neighbors, see Fig.~\ref{fig:mf}. Notice that in this case, we do not have absorbing states, but the plaquette terms are necessary to give origin to the chaotic oscillations.

By varying $J$, one can observe a bifurcation diagram of ``logistic'' type (period doubling) as  shown in Fig.~\ref{fig:mfb}. 

\begin{figure}
 \begin{center}
  \includegraphics[width=0.7\textwidth]{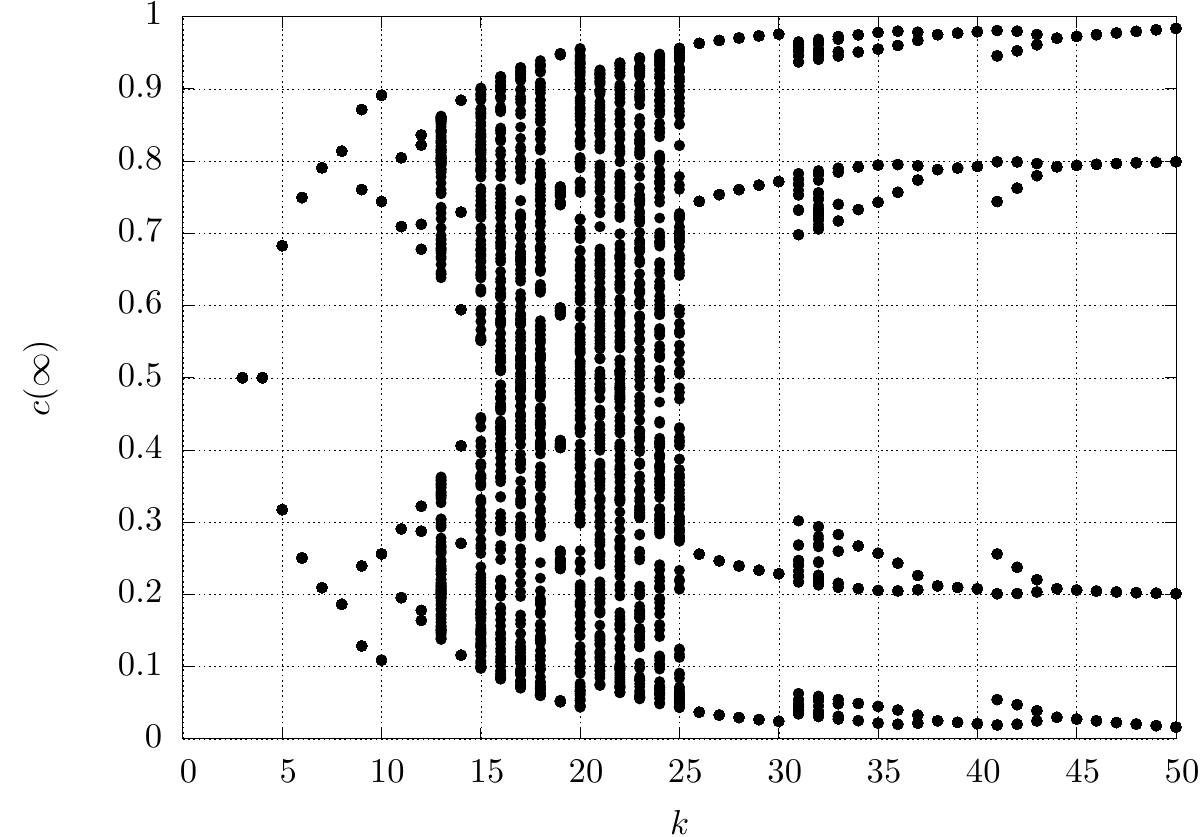}
  \caption{\label{fig:mfbifR} Mean field bifurcation diagram of Eq.~\eqref{mf} for $J=-6$ and  varying $k$, $q=0.1$, $\eps = 0.2$. Transient of 100 time steps, and plot of 10 iterations for 20 random initial conditions.}
 \end{center}
\end{figure}

The connectivity $k$ plays a fundamental role. As shown in Fig.~\ref{fig:mfbifR} for $J=-6$, chaotic oscillations occur only for some values of $k$. For $J=-6$, $k=20$ corresponds to an almost compact chaotic region, near to a window with a strongly periodic behavior.

Since we have approximated the behavior of an extended system with a scalar equation, we have imposed spatial homogeneity, which is not generally observed. Better approximations are obtained by replacing Eq.~\eqref{mf} with a spatial, coarse-grained description as follows
\begin{equation}\label{smf}
 c(t+1) = f(c) + D \De[2]{c}{x} + \eta \sqrt{c(1-c)},
\end{equation}
where $x$ represent the coarse-grained space index, $c = c(x,t)$, $D$ a diffusion coefficient, 
that also plays the role of a surface tension term that tends to make the system homogeneous. The term  $\eta \sqrt{c(1-c)}$  represents the local fluctuations of $c$. One can assume that it can be approximated by a white, delta-correlated noise term. This last term vanishes in correspondence of the absorbing states.

If the mean field part $f(c)$ converges towards an absorbing state, the dynamics is given by the competition between the diffusion and the noise term, and this produces the usual directed-percolation (or parity conservation)  phase transition, for which the presence of a stable, locally attracting absorbing state is essential~\cite{Hinrichsen}.
Far from the absorbing states (or if the absorbing states are not present, $\eps>0$) the noise term is irrelevant.

It is expected that a behavior more similar to that of mean-field can be observed if the diffusion term (or an equivalent mechanism) increases the spatial homogeneity. This is usually achieved in theoretical physics by increasing the dimensionality of the system, but this mechanism is unlikely to be observed in social networks.

In real society, people are rarely arranged in a one-dimensional lattice. There are many proposed structures for social networks, but one feature, the \emph{small-world effect} is generally present and its effect is that of an increasing spatial homogeneity.

\section{Smallworld bifurcations}

\begin{figure}
 \begin{center}
  \begin{tabular}{cc}
    $p=0$ & $p=0.1$\\[-.1cm]
    \includegraphics[width=0.4\textwidth]{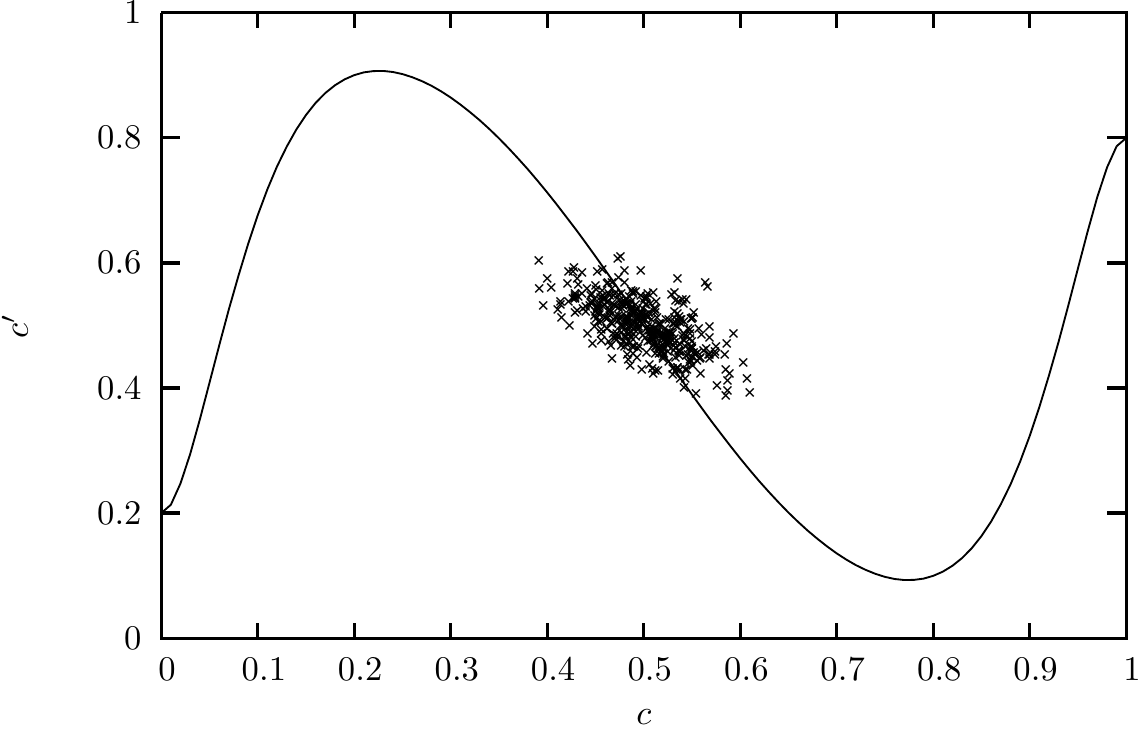} &
    \includegraphics[width=0.4\textwidth]{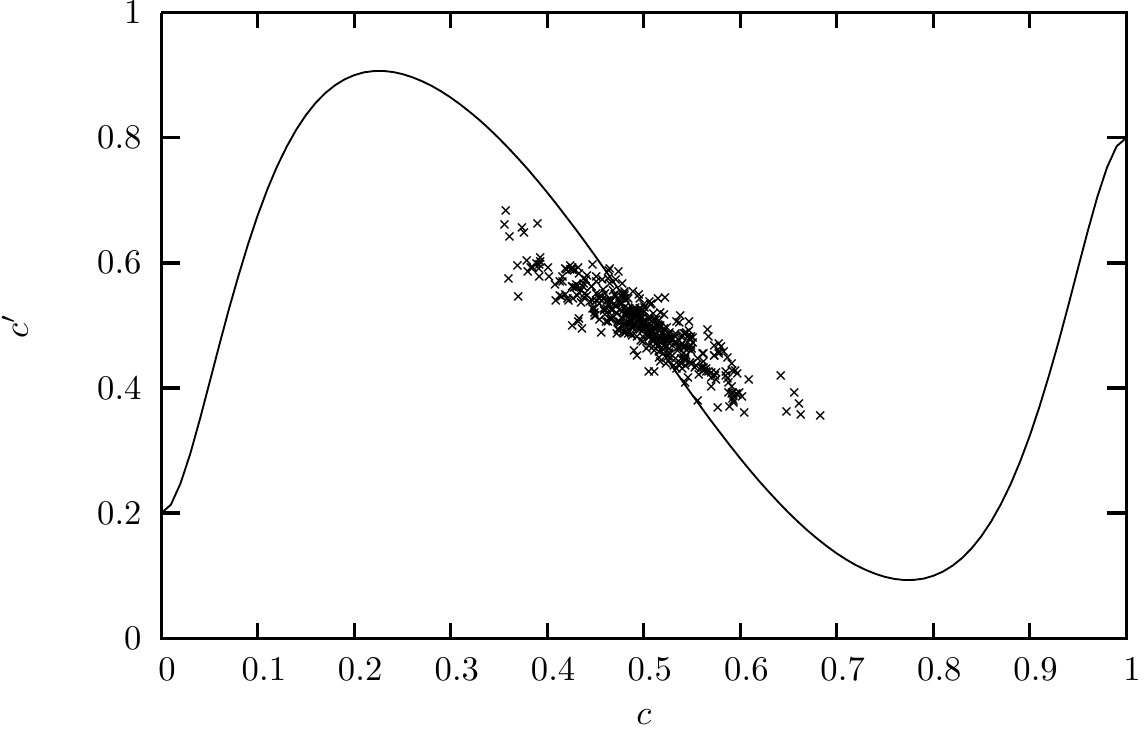} \\
    $p=0.25$ & $p=0.4$\\[-.1cm]
    \includegraphics[width=0.4\textwidth]{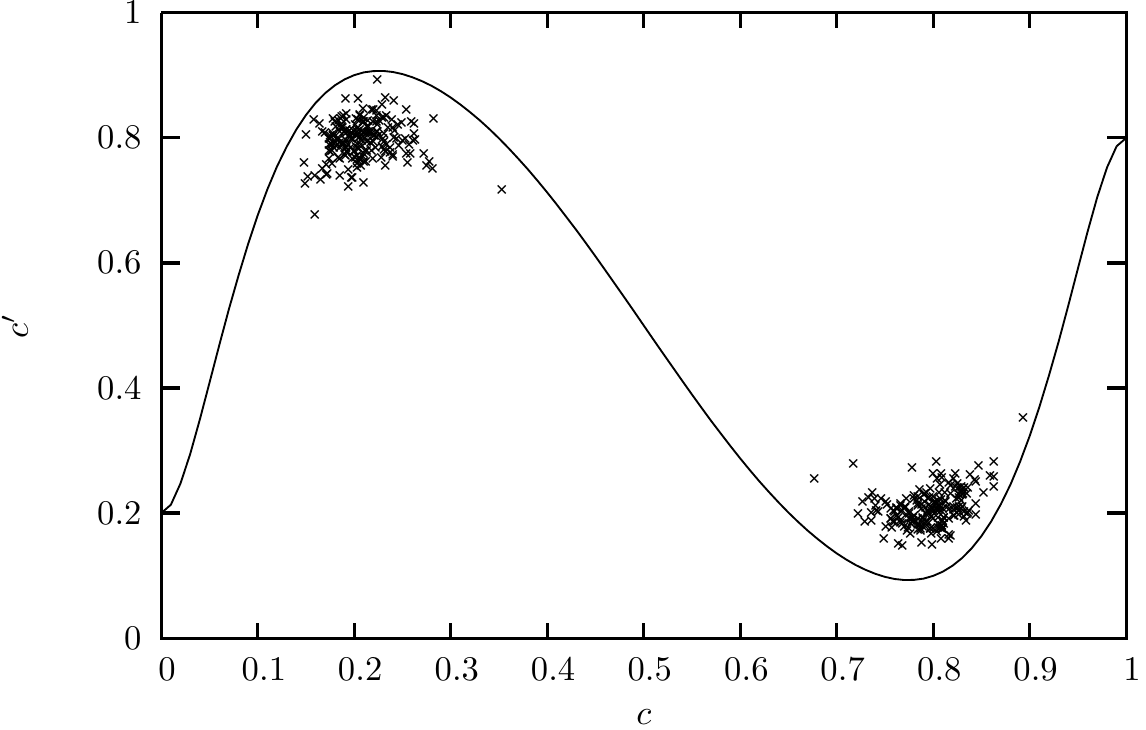} &
    \includegraphics[width=0.4\textwidth]{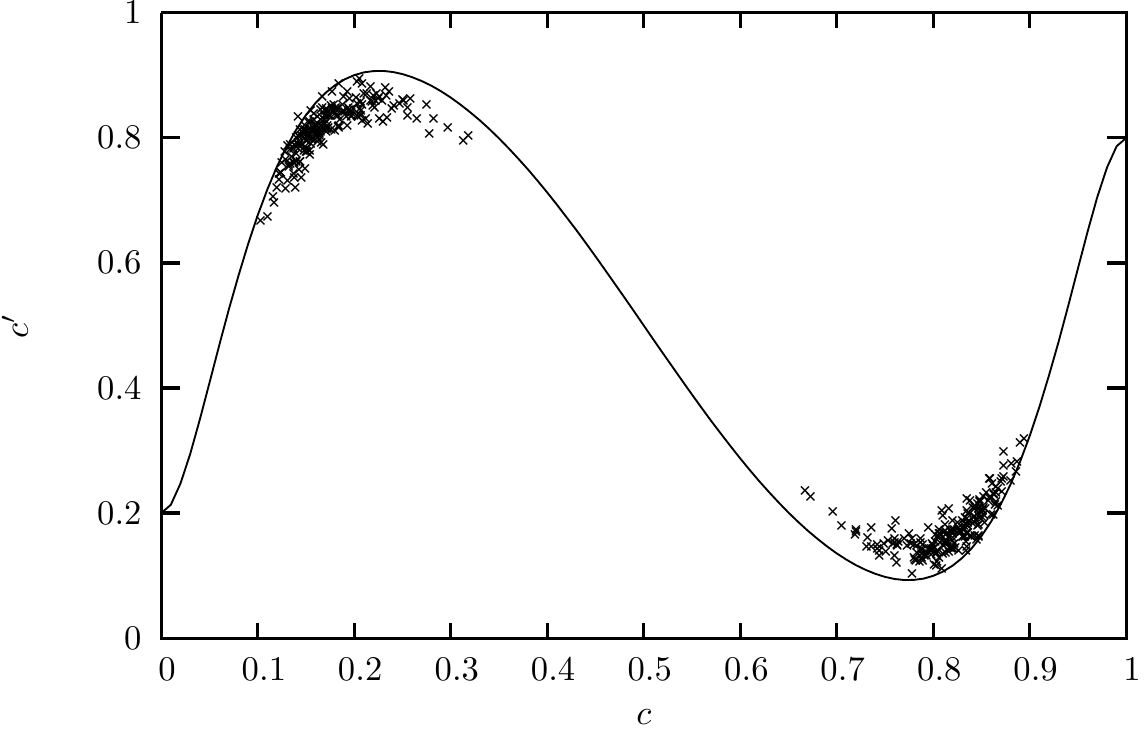} \\
    $p=0.5$ & $p=1$\\[-.1cm]
    \includegraphics[width=0.4\textwidth]{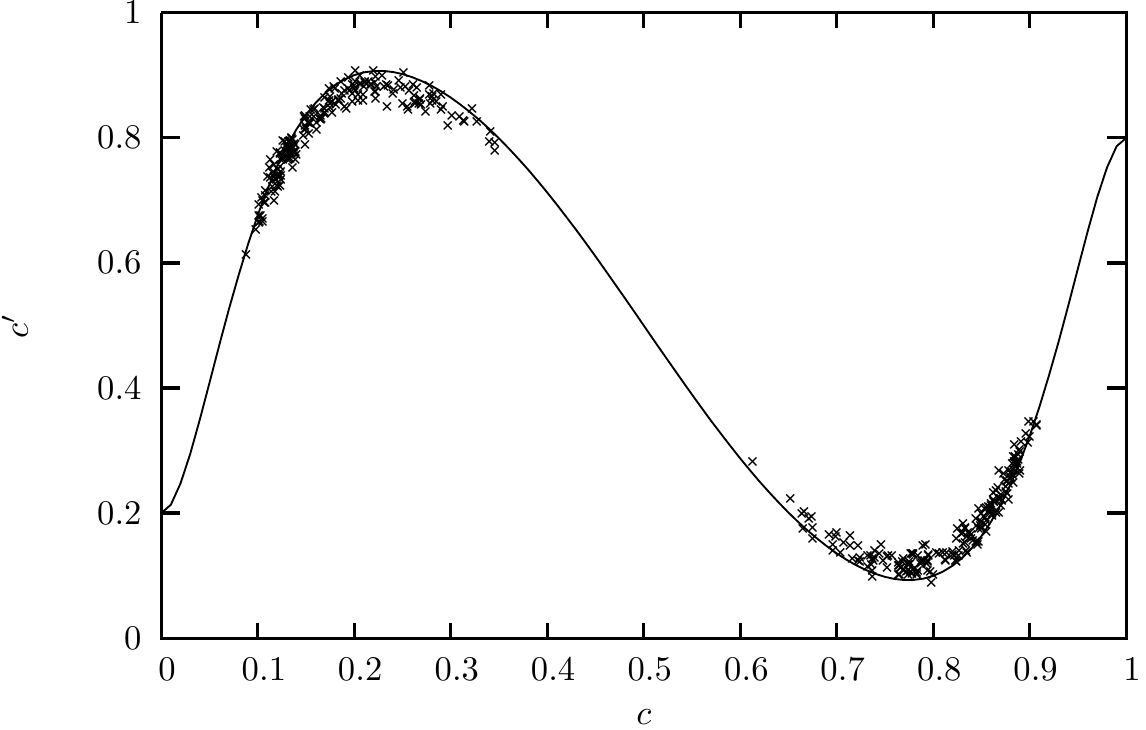} &
    \includegraphics[width=0.4\textwidth]{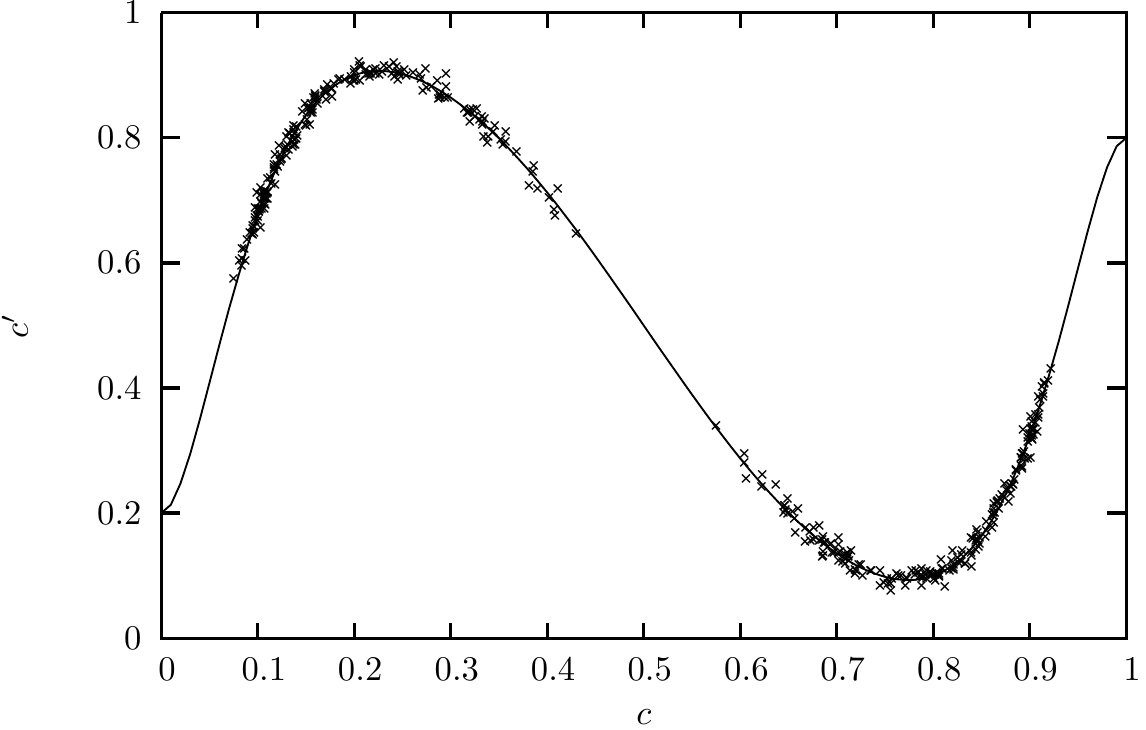}
  \end{tabular}
  \caption{\label{fig:conf} Density of opinion 1 ($c$) from microscopic simulations (crosses) plotted as return map [$c(t+1)$ vs. $c(t)$] and mean-field predictions for various values of the long-range connection probability $p$. Here $J=-3$, $k=20$, $q=0.1$, $\eps = 0.2$, $N=1000$, transient of 1000 time steps and plot of 200 iterations}
 \end{center}
\end{figure}

\begin{figure}
 \begin{center}
  \includegraphics[width=\textwidth]{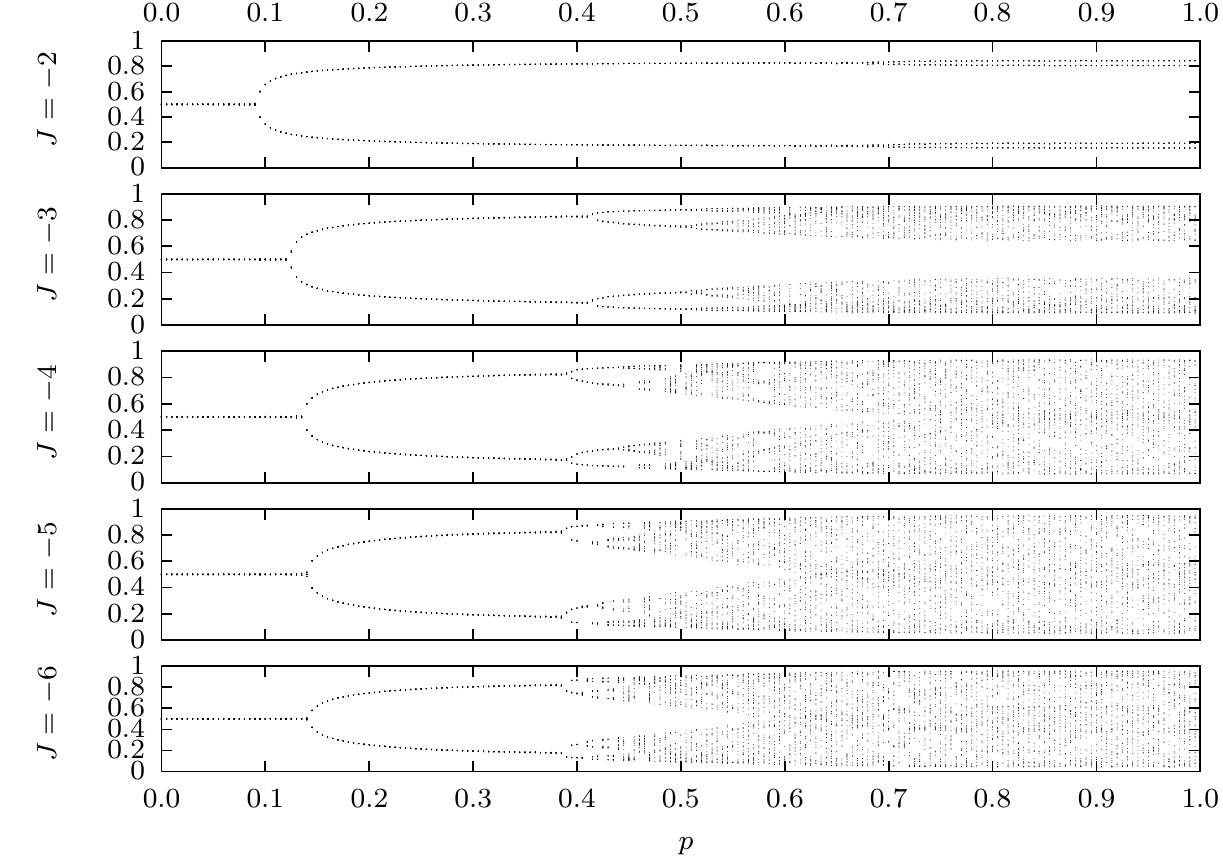}
  \caption{\label{fig:swb}  Small-world bifurcation diagram for $J=-2,\dots,-6$, $k=20$, $q=0.1$, $\eps = 0.2$, $N=5\cdot10^5$, transient of 2000 time steps and 2 different initial conditions plotted for 20 iterations.}
 \end{center}
\end{figure}

The Watts-Strogatz model~\cite{WattsStrogatz} is one of the simplest network models exhibiting the small-world effect which allows to smoothly change from a regular to a random lattice.
We have therefore simulated the microscopic system on a regular one-dimensional lattice where a fraction $p$ of links are rewired at random, and measured the behavior of the density $c(t)$. Its asymptotic value,  for large $t$, is denoted $c(\infty)$.

In Fig.~\ref{fig:conf}, we show the return map ($c(t+1)$ vs. $c(t)$) of the density of opinion 1 as results from the actual simulations, together with the mean-field predictions, for various values of $p$. One can see that for $p=0$, the density simply fluctuates around its mean value, 0.5. By increasing the fraction of long-range links, the distribution of points approaches more and more the mean-field predictions. 

We show in Fig.~\ref{fig:swb} the bifurcation diagram obtained by increasing the probability of long-range connections, $p$, for different values of $J$. 

By comparing the plots of Fig.~\ref{fig:mfb} and Fig.~\ref{fig:swb} one can see that one has close resemblance between the bifurcation induced by $J$ and that induced by $p$. However, we were not able to find an analytic mapping. Notice that this hypothetical map is nonlinear: the first bifurcation occurs at larger values of $p$ when decreasing $J$, but the second bifurcation moves in the opposite direction.

For $p$ above (roughly) 0.8, the distribution of points corresponds to that of mean-field, so one can assume that the small-world threshold for the chaotic oscillations of these models occurs at high values of the fraction of rewired links $p$, differently from the usual small-world effect. 

For $J>0$ the system is ``ferromagnetic'' (the transition probabilities are monotone with $c$, and so is $c'(c)$. Therefore there are at most three  fixed 
points corresponding to the intersection of the curve $c'(c)$ with the bisectrix (one for $c=0.5$ and two  symmetrically placed). By increasing $J$ a bifurcation appears (not reported here), 
very similar to that observed in an Ising model, but the branches 
corresponds to stable fixed points, not to cycles as for $J<0$. This scenario is confirmed by numerical simulations. For $p=0$, the system is ferromagnetic and finite-range, it is expected that this bifurcation is really stable only in the presence of absorbing states, \emph{i.e.}, for $\eps=0$, while for other values of $\eps$ it is only a metastable state (in real simulations). However, for $p > 0$ the system is actually infinite-range and therefore a phase transition can be present also in one-dimensional systems with finite interaction terms. It might be that a transition from metastable to stable states occurs for some finite values of $p$, even though we suspect that the threshold is for $p\rightarrow 0$. We have not investigated this aspect in detail.

\section{Conclusions}

We have studied an opinion model that exhibits, at the mean-field level, a period-doubling cascade to chaotic oscillations, by varying the coupling parameter $J$. The observed quantity is the average density of opinion 1, $c$.
Actual simulations on a one-dimensional lattice, in the absence of absorbing states or in the ``active'' phase, show \emph{microscopic} chaos, \emph{i.e.}, incoherent local oscillations; so that the density $c$ fluctuates around 0.5.

By rewiring a fraction $p$ of local connections to a random site, we trigger a small-world effect: the density $c$ exhibits a bifurcation diagram that resembles that obtained by varying $J$.
These small-world induced bifurcations are consistent with the general trend: long-range connections induce mean-field behavior. However, this is the first observation of such effect for a system exhibiting a \emph{chaotic} mean-field behavior. Indeed, the small-world effect makes the system coherent (with varying degree). We think that this observation may be useful since many theoretical studies of population behavior have been based on mean-field assumptions (differential equations), while actually one should rather consider individuals, and therefore spatially-extended, microscopic simulations. The well-stirred assumption is often not sustainable from the experimental point of view. However, it may well be that there is a small fraction of long-range interactions (or jumps), that might justify the small-world effect. In particular, it would be extremely useful to derive a general ``rescaling'' formulation allowing the estimation of the \emph{effective} value of parameters given a certain degree of small-world connections.

For what concerns our specific opinion model, we can draw some sociological consequences from our simplified assumptions. We simulated social groups with ``frustrated'' behavior, \emph{i.e.}, conformistic for a strong social pressure and anticonformistic for ``marginal'' behavior like fashion or dressing. Such a scheme can be probably applied to many societies in the transition phase from traditional to non-traditional behavior, but also to social micro-cosmos in western societies, in which social norms are hardly broken. It is plausible that these frustrations may trigger oscillations, possibly chaotic. A social initiative promoting homogenization, or the social mixing due to living or working conditions, could act by favoring long-range interactions and triggering coherent oscillations. Such oscillations could be identified in the sudden explosion of violence or pathological trends (say suicide, self-mutilations, etc.).

\end{document}